\documentclass[sigconf,nonacm,screen]{acmart}
\usepackage{booktabs,tabularx,tablefootnote}
\usepackage{colortbl}

\setcitestyle{authoryear,open={(},close={)}}

\AtBeginDocument{%
  \providecommand\BibTeX{{%
    \normalfont B\kern-0.5em{\scshape i\kern-0.25em b}\kern-0.8em\TeX}}}

\setcopyright{acmcopyright}
\copyrightyear{2024}
\acmYear{2024}
\acmDOI{XXXXXXX.XXXXXXX}

\acmConference[Name '24]{Make sure to enter the correct conference title from your rights confirmation email}{February 27--March 3, 2024}{}
\acmPrice{15.00}
\acmISBN{978-1-4503-XXXX-X/18/06}

\begin{document}

\title{Authorship Attribution in the Era of LLMs:\\ Problems, Methodologies, and Challenges}

\author{Baixiang Huang}
\email{baixiang.huang@emory.edu}
\affiliation{
  \institution{Emory University}
  \city{Atlanta}
  \state{GA}
  \country{USA}
}

\author{Canyu Chen}
\email{cchen151@hawk.iit.edu}
\affiliation{
  \institution{Illinois Institute of Technology}
  \city{Chicago}
  \state{IL}
  \country{USA}
}

\author{Kai Shu}
\email{kai.shu@emory.edu}
\affiliation{
  \institution{Emory University}
  \city{Atlanta}
  \state{GA}
  \country{USA}
}

\renewcommand{\shorttitle}{Authorship Attribution in the Era of LLMs: Problems, Methodologies, and Challenges}

\begin{abstract}
Accurate attribution of authorship is crucial for maintaining the integrity of digital content, improving forensic investigations, and mitigating the risks of misinformation and plagiarism. Addressing the imperative need for proper authorship attribution is essential to uphold the credibility and accountability of authentic authorship. The rapid advancements of Large Language Models (LLMs) have blurred the lines between human and machine authorship, posing significant challenges for traditional methods. We present a comprehensive literature review that examines the latest research on authorship attribution in the era of LLMs. This survey systematically explores the landscape of this field by categorizing four representative problems: (1) Human-written Text Attribution; (2) LLM-generated Text Detection; (3) LLM-generated Text Attribution; and (4) Human-LLM Co-authored Text Attribution. We also discuss the challenges related to ensuring the generalization and explainability of authorship attribution methods. Generalization requires the ability to generalize across various domains, while explainability emphasizes providing transparent and understandable insights into the decisions made by these models. By evaluating the strengths and limitations of existing methods and benchmarks, we identify key open problems and future research directions in this field. This literature review serves as a roadmap for researchers and practitioners interested in understanding the state of the art in this rapidly evolving field. Additional resources and a curated list of papers are available and regularly updated at \href{https://llm-authorship.github.io/}{\color{ACMPurple}https://llm-authorship.github.io}.
\end{abstract}

\maketitle

\section{Introduction}
Authorship Attribution (AA) is the process of determining the author of a particular piece of writing, which has significant real-world applications across various domains. In forensic investigations, authorship attribution plays a crucial role in solving murder cases disguised as suicides \cite{chaski2005forensic_court,grant2020forensic_murder}, tracking terrorist threats \cite{winter2019terrorism,cafiero2023aa_qanon}, and aiding general criminal investigations \cite{koppel2008law_enforcement,argamon2018forensic,belvisi2020forensic}. In the digital realm, authorship attribution helps safeguard the integrity of content by preventing deceptive social media activities \cite{hazell2023phising}, detecting account compromises \cite{barbon2017social_media}, and linking user profiles across various social networks \cite{shu2017link_user,sinnott2021social_mdeia}. Additionally, authorship attribution techniques are instrumental in combating misinformation \cite{chen2024combatingmisinformation,chen2024can,shu2020misinfo,stiff2022disinfo,hanley2024misinfo}, protecting intellectual property rights \cite{meyer2007plagiarism,stamatatos2011plagiarism}, and identifying fraudulent practices such as fake reviews \cite{ott2011data_opinion_spam,afroz2012hoax_fraud}.

\begin{figure}[t]
    \centering
    \includegraphics[width=0.48\textwidth]{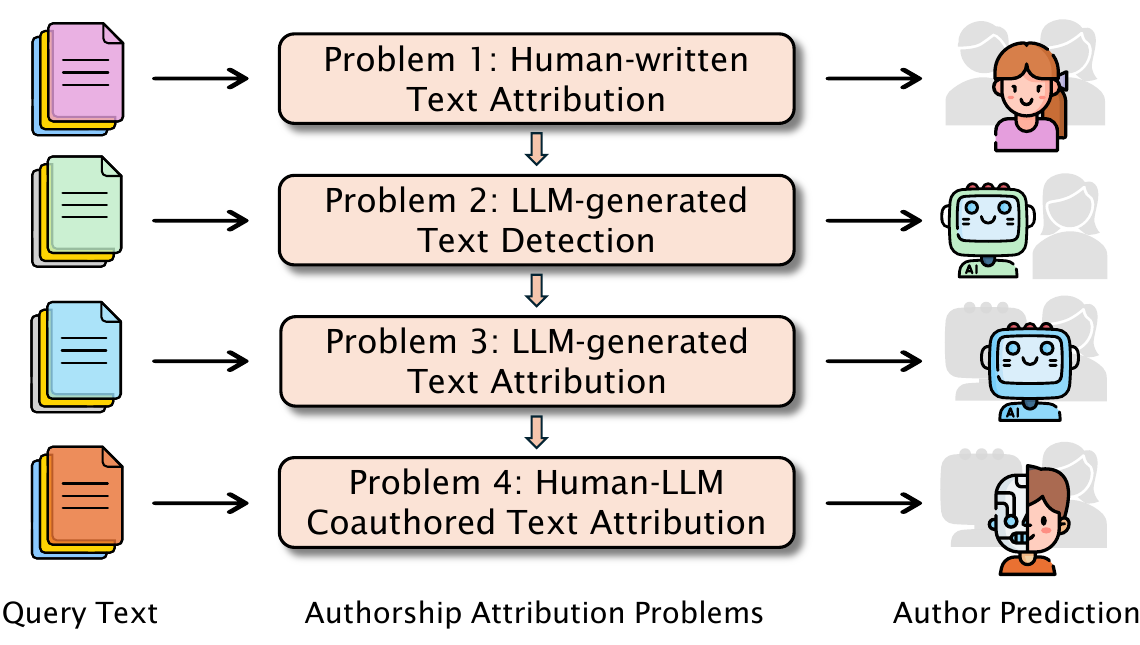}
    \caption{Four representative Problems in Authorship Attribution: (1) Human-written Text Attribution, which involves attributing an unknown text to its human authors; (2) LLM-generated Text Detection, which focuses on detecting whether a text has been generated by LLMs; (3) LLM-generated Text Attribution, aimed at identifying the specific LLM or human responsible for a given text; (4) Human-LLM Co-authored Text Attribution, which classifies a text as human-written, LLM-generated, or a combination of both. \textit{These problems become progressively more complex, as indicated by the arrows.}} 
    \vspace{-0.5cm}
    \label{fig:intro}
\end{figure}

The development of large language models (LLMs) has revolutionized text generation, offering numerous benefits but also raising significant concerns about text authenticity and originality \cite{goldstein2023generative_model_misuse}. The advent of LLMs has complicated authorship attribution, making it increasingly difficult to distinguish between LLM-generated texts and human-written texts \cite{clark2021ntg_evaluation,sadasivan2023_machine_txt}. Identifying LLM-generated texts is challenging even for human experts, let alone traditional authorship attribution methods \cite{liu2024chatgpt_academic_gpabenchmark,gao2022chatgpt_detect}. This inability to distinguish between human and machine-generated content undermines the integrity of authorship, complicates legal and ethical responsibilities, and threatens the credibility of digital content and the safety of online space~\cite{solaiman2023evaluating,vidgen2024introducing}. 

Over the past few decades, authorship attribution has experienced significant advancements due to the development of natural language analysis tools and innovative methods for text representation learning. Traditionally, authorship attribution relied on stylometry, which analyzes an individual's unique writing style through feature engineering to capture linguistic characteristics \cite{lagutina2019survey_stylometry}. The emergence of machine learning algorithms capable of handling high-dimensional data has enabled the creation of more expressive representations. In recent years, there has been a shift towards extracting text embeddings using pre-trained language models \cite{fabien2020aa_bert_classifiy}. These approaches, while offering higher performance, often sacrifice explainability for accuracy \cite{rivera2021luar}. More recently, researchers have begun utilizing LLMs to extract features in conjunction with machine learning classifiers or to conduct end-to-end reasoning for authorship attribution \cite{patel2023lisa,huang2024authorship}. 

The rapid advancements in LLMs have significantly improved text generation, producing outputs that rival human writing in fluency and coherence. This progress underscores the imperative need to distinguish between human-written text, LLM-generated text, or a combination of both. As illustrated in Figure \ref{fig:intro}, authorship attribution can be systematically categorized into four representative problems: attributing unknown texts to human authors, detecting LLM-generated texts, identifying the specific LLM or human responsible for a text, and classifying texts as human, machine, or human-LLM co-authored. Each task presents unique challenges that necessitate corresponding solutions. Researchers continually adapt and refine attribution methods, transitioning from human-authored texts to LLM-generated content, and navigating the complex interweaving in human-LLM co-authored works. As detection methods advance, adversarial attacks also evolve to bypass these measures, creating a continuous cycle of challenge and response in the quest to distinguish and disguise authorship \cite{dugan2024raid}. Addressing these challenges will pave the way for more robust and reliable authorship attribution techniques.

Authorship attribution for both human and LLM-generated texts can be framed as either binary or multi-class classification. The LLM-generated text detection task simplifies the attribution problem by classifying each text as either originating from humans or LLMs \cite{jawahar2020survey_machine,mitchell2023detect_gpt,pu2023machine_detect,sadasivan2023_machine_txt}. The majority of previous research on the automatic detection of machine-generated text has focused on binary classification \cite{jawahar2020survey_machine,mitchell2023detect_gpt}. In the more challenging multi-class task, the goal is not only to differentiate between human and LLM-generated text but also to classify the text according to its specific source of generative models \cite{uchendu2021bench_turing,li2023sniffer_origin_tracing}. Differences in LLM architectures, training methods, and generation techniques can influence the style of the generated texts \cite{munir2021aa_machine}. In the more complex human-LLM co-authoring problem, the goal is to distinguish texts authored by humans, LLMs, or combinations of both. Such nuanced detection provides deeper insights into the provenance of the text and is crucial for applications requiring detailed source attribution. Neural network-based detectors, generally outperform metric-based methods in both human authorship attribution and LLM-generated text detection problems \cite{he2023mgtbench,zhang2024mixset}. However, these neural network approaches often offer less explainability compared to their metric-based counterparts.

This review serves as a valuable resource, comprehensively summarizing existing literature and highlighting the challenges and opportunities introduced by LLMs. We provide an in-depth analysis of methodologies and datasets in this evolving field. The main contributions of this paper are as follows:
\begin{itemize}
    \item We provide a timely overview to discuss the challenges and opportunities presented by LLMs in the field of authorship attribution. By systematically categorizing authorship attribution into four representative problems and balancing problem complexity with practicality, we reveal insights into the evolving field of authorship attribution.
    
    \item We offer a comprehensive comparison of state-of-the-art methodologies, datasets, benchmarks, and commercial tools used in authorship attribution. This analysis not only improves the understanding of authorship attribution but also provides a valuable resource for researchers and practitioners to use as guidelines for approaching this direction.
    
    \item We discuss open issues and provide future directions by considering crucial aspects such as generalization, explainability, and interdisciplinary perspectives. We also discuss the broader implications of authorship attribution in real-world applications. This holistic approach ensures that authorship attribution not only yields accurate results but also provides insights that are explainable and socially relevant.
\end{itemize}

The remainder of this survey is organized as follows. In Section \ref{p1}, we examine human authorship attribution, starting with a problem definition and moving through various methodologies, followed by a discussion on challenges. In Section \ref{p2}, we discuss LLM-generated text detection. In Section \ref{p3}, we explore LLM-generated text attribution. Section \ref{p4} covers Human-LLM co-authored text attribution. Section  \ref{resources} discusses resources and evaluation metrics, comparing benchmarks and datasets. Section \ref{opportunities} highlights opportunities and future directions. In Section \ref{ethic}, we discuss ethical and privacy concerns. Finally, we conclude this survey in Section \ref{conclusion}.

\section{Human Authorship Attribution} \label{p1}
This section discusses authorship attribution of human-written texts. It covers various methodologies, including stylometry, machine learning, pre-trained language models, and LLM-based methods. Challenges such as limited data, evolving writing styles, and interpretability are also addressed.

\subsection{Problem Definition}
Authorship attribution aims to identify the author of an unknown text from a set of known authors. This can be an open-class problem, where the true author might not be among the known authors, or a closed-class problem, where the true author is included in a finite set of authors \cite{stolerman2014attribution_open,andrews2019conv_trans}. Authorship attribution methods are typically divided into classification-based methods for a small set of candidate authors and similarity-based ranking methods for larger numbers of authors \cite{rivera2021luar,huertas2022part}. These techniques can also be adapted to related problems such as authorship verification and profiling. Authorship verification determines whether a piece of writing was written by a specific individual \cite{stamatatos2016review_av}, while profiling infers characteristics such as age or gender from the author's writing style \cite{argamon2009profiling_bayes}.

\subsection{Methodologies}
We explore the evolution of methods used to analyze human-written text, starting with stylometry. Over time, the focus has shifted to using text embeddings. Recently, the integration of LLMs marks further advancements in attributing authorship.

\paragraph{Stylometry Methods}
Stylometry, the quantitative analysis of writing style, has evolved from its initial reliance on human expertise \cite{mosteller1963aa_bayes} to computational methods \cite{neal2017survey_stylometry,lagutina2019survey_stylometry}. This discipline utilizes a variety of linguistic features to determine authorship \cite{holmes1994survey_aa,lagutina2019survey_stylometry}, positing that each author's unique style can be captured through quantifiable characteristics \cite{argamon2009profiling_bayes}. Key stylometric features include character and word frequencies \cite{sharma2018aa_ngram}, parts of speech \cite{sundararajan2018aa_pos}, punctuation, topics \cite{seroussi2014data_imdb_aa_topic_model,potha2019av_topic_model,halvani2021av_mask_topic}, and vocabulary richness. Important features can be categorized into the following types: lexical, syntactic, semantic, structural, and content-specific \cite{rudman1997survey_aa_problems}. Lexical features involve word choice and frequency; syntactic features pertain to sentence structure and grammar; semantic features explore the meaning and context of words; structural features relate to text organization; and content-specific features emphasize domain-specific terms \cite{bozkurt2007aa,seroussi2014data_imdb_aa_topic_model}.

\paragraph{Machine Learning Methods}
Machine learning approaches integrate stylometric features with classifiers such as logistic regression \cite{aborisade2018tweet_logistic_regression,madigan2005aa_lr}, Bayesian multinomial regression \cite{argamon2009profiling_bayes}, and support vector machines (SVM) \cite{bacciu2019aa_cross_domain_multi_lang_ensemble_svm}. Before the widespread adoption of transformer-based models, multi-headed Recurrent Neural Networks (RNNs) \cite{bagnall2015aa_rnn}, and Long Short-Term Memory (LSTMs) were utilized at both sentence and article levels \cite{qian2017aa_lstm}. Convolutional Neural Networks (CNNs) were applied at various levels, including characters, words, and N-grams \cite{ruder2016aa_cnn,shrestha2017aa_cnn_n_gram_tweet}. Moreover, syntax-augmented CNN models \cite{zhang2018aa_syntax_cnn}, Convolutional Siamese Networks \cite{saedi2021aa_cnn_siamese}, and attention-based Siamese Networks \cite{boenninghoff2019explain_siamese_lstm} were explored. Additionally, combinations of convolutions and transformers have been employed to learn embeddings for comparison tasks \cite{andrews2019conv_trans}.

\paragraph{Pre-trained Language Models}
Pre-trained language models, especially BERT-based architectures \cite{devlin2018bert} such as BERT \cite{ippolito2019detect_bert,fabien2020aa_bert_classifiy,manolache2021av_bert}, Sentence-BERT \cite{schlicht2021sbert_hate_speech,rivera2021luar}, and RoBERTa \cite{huertas2022part}, have proven effective for learning authorship representation. These methods do not require hand-crafted features but require substantial training time and domain-specific labeled data, struggling with cross-domain generalization and explainability. Contrastive learning \cite{khosla2020contrastive_learning} is often used with pre-trained language models to enhance stylistic representation by maximizing similarity between texts written by the same authors and minimizing it with texts from different authors \cite{huertas2022part}. 

\citet{barlas2020aa_cross_domain_bert} found that BERT performed well with large vocabularies, outperforming multi-headed RNNs. \citet{fabien2020aa_bert_classifiy} fine-tuned a BERT model for authorship attribution, showing that including additional stylometric and hybrid features in an ensemble model can improve performance. \citet{rivera2021luar} concluded that topic diversity and dataset size are crucial for effective cross-domain transfer. Adaptation through style transfer has not resolved cross-domain issues \cite{boenninghoff2019av,wegmann2022av_transfer}. Techniques like slanted triangular learning rates and gradual unfreezing can be used to avoid catastrophic forgetting during fine-tuning \cite{howard2018lm_fine_tune}.

\paragraph{LLM-based Methods}
Despite advances in LLMs, their potential for authorship attribution remains underexplored. The natural language understanding capability allows LLMs to recognize nuances, styles, and patterns in language, which are crucial for distinguishing between authors. Authorship attribution is a complex reasoning task, and LLMs possess significant capabilities in reasoning and problem-solving, particularly in zero-shot learning within resource-limited domains \cite{kojima2022zero_shot_reasoners}. They assist in feature extraction by identifying syntactic patterns, lexical choices, and grammatical structures that are essential for authorship attribution. Traditionally, LLMs have been employed mainly for auxiliary tasks such as feature extraction and data annotation \cite{patel2023lisa}. Notable examples include the use of GPT-3 for data annotation \cite{brown2020gpt3} and a T5 Encoder for learning authorship signatures. Beyond feature extraction, LLMs can also be directly utilized for end-to-end authorship attribution \cite{huang2024authorship}.

The incorporation of LLMs in authorship attribution addresses several limitations of traditional methods. Unlike BERT-based models, which require computationally expensive fine-tuning and large amounts of domain-specific data for optimal performance, LLMs can generalize across various domains without fine-tuning, thereby mitigating issues related to domain specificity \cite{barlas2020aa_cross_domain_bert}. LLMs are also effective with shorter texts, reducing the necessity for long inputs to derive meaningful representations \cite{eder2015aa_text_size}. Another key advantage of LLM-based approaches is their ability to provide natural language explanations for their predictions, enhancing transparency compared to hidden text embeddings \cite{huang2024authorship}. This versatility marks a step forward in overcoming challenges related to data, domain specificity, text length requirements, and explainability faced by earlier methods.

\subsection{Open Challenges}
Human authors exhibit a diverse range of writing styles influenced by genre, topic, context, and temporal changes. This variability complicates authorship attribution, necessitating the identification of consistent and unique stylistic markers. Further complexity arises from the presence of noise due to varying document sizes and languages, requiring algorithms to manage a broad spectrum of linguistic nuances. Short or low-quality texts, such as social media posts, often provide unreliable data, complicating accurate attribution \cite{eder2015aa_text_size,theophilo2021social_msg_short}. Collaborative texts, which blend multiple writing styles, further mask individual contributions and obscure distinct authorial signals \cite{dauber2017stylometric_collab}. 

Traditional stylometric methods rely on human expertise and manually crafted features, whereas deep learning methods demand significant computational resources and extensive labeled data, with the risk of catastrophic forgetting \cite{ramasesh2021effect_forgetting}. Authorship attribution using LLMs also faces several challenges: their effectiveness decreases as the number of candidate authors increases due to context length constraints \cite{huang2024authorship}, and they can perpetuate biases from training data, resulting in inaccurate attributions for texts from marginalized groups and languages \cite{liang2023bias_non_en}. Additionally, LLMs can be misused to generate content that conceals true authorship by mimicking others or using LLMs to alter their work.

\section{LLM-generated Text Detection} \label{p2}
LLMs excel at generating fluent and coherent text, raising concerns about the authenticity and originality of authorship. Detecting LLM-generated text is crucial for several applications, including combating misinformation and disinformation on social media \cite{gambini2022ntg_tweet,stiff2022disinfo,chen2024combatingmisinformation}, identifying spam \cite{jindal2008opinion_spam}, preventing phishing attacks \cite{hazell2023phising}, identifying fake reviews \cite{salminen2022fake_reviews}, detecting machine-generated scientific papers \cite{rodriguez2022cross_academia,liu2024chatgpt_academic_gpabenchmark}, and making academic peer review judgments \cite{li2025llm_judgement}. As a result, the detection of LLM-generated text has garnered significant attention \cite{kumarage2023aa_llm_neural_text,tang2023llm_detect,wu2023survey_llm_text,yang2023survey_llm_text}.

\subsection{Problem Definition}
LLM-generated texts are included within the scope of machine-generated texts\footnote{Machine-generated texts is also referred to as machine-authored, AI-generated, neural-generated, deepfake text, neural text, or synthetic text.}. Machine-generated texts encompass any text produced by automated systems, including simpler language models or rule-based systems \cite{uchendu2021bench_turing}. This paper focuses specifically on LLM-generated texts. The task of detecting LLM-generated text involves identifying and distinguishing text created by LLMs from that written by humans. Typically, this task is approached as a binary classification problem \cite{zellers2019data_grover,solaiman2019tf_idf_logistic,jawahar2020survey_machine,fagni2021tweepfake,mitchell2023detect_gpt}.

\subsection{Methodologies}
Evaluating the quality of machine-generated excerpts has traditionally relied on human judgment, which is considered the gold standard for open-domain generation systems \cite{van2019eval_human,gehrmann2019machine_text_statisc}. However, distinguishing between LLM-generated and human-written texts poses significant challenges for humans \cite{dugan2023eval_human}. For example, untrained human reviewers often fail to distinguish GPT-3-generated text from human-written text, identifying it correctly only at a rate consistent with random chance \cite{clark2021ntg_evaluation}. \citet{liu2024chatgpt_academic_gpabenchmark} found that even experienced faculty members and researchers could only achieve about a 50\% success rate in identifying GPT-generated academic writings. In contrast, detection algorithms frequently outperform human reviewers in this task \cite{ippolito2019detect_bert}.

\citet{chakraborty2023possibilities} employed theoretical analysis to argue that detecting LLM-generated text is nearly always feasible with the collection of multiple samples, and they established precise sample complexity bounds for this detection. However, existing detectors and models for LLM-generated text are not yet fully reliable \cite{sadasivan2023_machine_txt, wang2023bench_m4, dugan2024raid}. \citet{sadasivan2023_machine_txt} provided theoretical insights indicating that the detection problem is becoming increasingly difficult.

LLM-generated text detectors can be categorized into metric-based and model-based methods \cite{he2023mgtbench}, which are further divided into feature-based, neural network-based, zero-shot-based, and watermark-based methods. These detectors are also classified as either white-box or black-box, depending on their access to LLM weights \cite{tang2023llm_detect,yang2023survey_llm_text}. Watermarking-based methods typically fall under white-box detection, while proprietary models are restricted to black-box methods.

\paragraph{Featured-based Method}
LLM-generated texts are typically less emotional and more objective than human-written texts \cite{guo2023data_hc3}. In comparison, human-authored texts are generally more coherent, while LLM-generated texts tend to repeat terms within a paragraph \cite{dugan2023eval_human}. Similar to the problem of human authorship attribution, linguistic features such as phrasal verbs, coreference, part-of-speech (POS) tags, and named entity (NE) tags are also useful in distinguishing LLM-generated text \cite{nguyen2017machine_feature,see2019storyteller,frohling2021machine_feature}. Feature-based methods are more explainable, but have drawbacks, such as poor generalizability of certain features across different domains and sampling methods.

\paragraph{Neural Network-Based Detectors}
Neural network-based detectors, particularly those utilizing BERT, have proven effective in distinguishing between human-written texts and those generated by GPT-2 \cite{ippolito2019detect_bert,liu2019roberta}. \citet{solaiman2019tf_idf_logistic} fine-tuned the RoBERTa model with a dataset of GPT-2 outputs in open domain settings. Similarly, \citet{guo2023data_hc3} fine-tuned RoBERTa to detect ChatGPT-generated text. \cite{zhan2023g3detector} developed G3Detector by fine-tuning RoBERTa-large for the same purpose. Additionally, \citet{chen2023gpt_sentinel} introduced GPT-Sentinel, training both RoBERTa and T5 \cite{raffel2020t5} on their OpenGPTText dataset. In a different approach, \citet{hu2023radar_vicuna} created RADAR, which fine-tunes Vicuna 7B \cite{vicuna2023} in a generative adversarial setting along with a paraphrase model. These efforts highlight the ongoing advancements in enhancing the detection of LLM-generated content using BERT-based models.

These detectors require retraining when encountering text from new LLMs to ensure reliable detection \cite{mitchell2023detect_gpt,chakraborty2023robust}. Neural network-based detectors are vulnerable to adversarial and poisoning attacks \cite{goodfellow2014adversarial,wang2022poisoning,pu2023machine_detect}. These detectors also face limitations such as overfitting to training data \cite{uchendu2020aa_neural_liwc}. The generalization ability of these detectors is critical, having been trained on various model families and tested on unseen models \cite{pu2023generalization,bhattacharjee2023conda_generalization}. Surrogate models, which are often small language models, are also applied to train classifiers \cite{verma2023ghostbuster,mireshghallah2023smaller}.

\paragraph{Zero-Shot Detectors}
Zero-shot detection methods are generally statistics-based, enabling the identification of LLM-generated text without additional training \cite{su2023detectllm}. Various statistical measures have been employed for detection, including entropy \cite{lavergne2008machine_text_entropy,gehrmann2019machine_text_statisc}, perplexity \cite{beresneva2016machine_text_perplexity,hans2024binocular}, average log-probability score \cite{solaiman2019tf_idf_logistic}, fluency \cite{holtzman2019neural_text}, and Zipf's word frequency law \cite{zipf2016word_freq,piantadosi2014zipf}\footnote{The frequency of a word decreases as its rank in a frequency-ordered list increases.}. Additional methods leverage n-grams \cite{yang2023dna_gpt}, Uniform Information Density (UID) \cite{venkatraman2023gptwho}, log rank information \cite{su2023detectllm}, and various linguistic features such as part-of-speech determiners, conjunctions, auxiliary relations, vocabulary, and emotional tone \cite{joulin2016fast_text,tang2023llm_detect,gehrmann2019machine_text_statisc}.

Various zero-shot detection methods provide distinct strategies to enhance both detection accuracy and efficiency. \citet{galle2021unsupervised} introduced an unsupervised method that identifies the over-appearance of repeated higher-order n-grams, distinguishing them from human-generated text. DetectGPT \cite{mitchell2023detect_gpt} relies on the observation that LLM-generated passages often fall into regions of negative curvature in log probability. \cite{bao2023fast_detectgpt} enhanced this approach and proposed Fast-DetectGPT, which increases efficiency by using conditional probability curvature. Additionally, some methods leverage LLMs themselves for text classification \cite{zellers2019data_grover,solaiman2019tf_idf_logistic}. Different decoding methods are often applied to generate more diverse and less repetitive text, though these can also lead to hallucinations and less verifiable content \cite{shakeel2021fake_news_fact,guo2023data_hc3}. Fact-checking methods can mitigate these issues \cite{zhong2020fact_check,schuster2020fake_news_fact_check}. Additionally, \citet{krishna2024adv_paraphrasing_retrieval} developed a detector that uses information retrieval to store LLM outputs in a database and search for semantically similar content to identify LLM-generated text, though this method raises privacy concerns regarding the storage of user conversations.

\paragraph{Watermarking}
Watermarking involves embedding specific patterns within text generated by LLMs, making them imperceptible to humans but detectable through specialized methods. \cite{topkara2005watermark,meral2009watermark,kirchenbauer2023watermark_llm,zhao2023watermark_robust}. By imprinting distinct patterns, watermarking enables the identification of LLM-generated text. Various methods include parsed syntactic tree structures \cite{atallah2001watermark_syntactic_tree,topkara2005watermark}, synonym tables \cite{jalil2009watermark_synonym_table}, adversarial watermarking \cite{abdelnabi2021adversarial}, and context-aware lexical substitution \cite{yang2022watermark_lexical_substitution}. One notable approach is soft watermarking, proposed by \citet{kirchenbauer2023watermark_llm}, which partitions tokens into “green” and “red” lists to create patterns. A watermarked LLM samples tokens from the green list with high probability, determined by a pseudo-random generator seeded by its prefix token. The watermarking detector classifies passages with a high frequency of tokens from the green list as LLM-generated.

Further advancements have improved the robustness, efficiency, and stealthiness of watermarking methods \cite{hou2023watermark_robust,wu2023watermark_resilient,zhao2023watermark_robust}. However, there is a trade-off between watermark effectiveness and text quality, as more reliable watermarks require more extensive text modifications \cite{sadasivan2023_machine_txt}. Additionally, watermarking presents challenges for proprietary LLMs and third-party applications due to the necessity of accessing the language model's logits \cite{kirchenbauer2023watermark_llm}. Watermark-based detection methodologies are also vulnerable to paraphrasing attacks \cite{sadasivan2023_machine_txt,krishna2024adv_paraphrasing_retrieval}.

\subsection{Open Challenges}
Detecting LLM-generated texts is challenging due to their high quality and ability to mimic human writing styles. LLMs can emulate human writing so closely that they pose significant challenges to traditional stylometric techniques. They can incorporate complex narrative structures and varied vocabularies, making it difficult to distinguish between human and LLM-generated texts. The rapid evolution of LLMs further complicates detection, as newer versions exhibit different stylistic traits, making detection models quickly obsolete \cite{chakraborty2023robust}. LLM-generated text detectors often struggle to generalize to unseen domains encountered during training \cite{pu2023generalization,rodriguez2022cross_academia,li2024detect_mage} and tend to perform better on LLMs they were specifically trained on \cite{pu2023generalization,chakraborty2023robust,li2024detect_mage}. 

Existing detectors also lack robustness to various factors, such as alternative decoding strategies \cite{ippolito2019detect_bert}, input sequence length \cite{solaiman2019tf_idf_logistic}, different prompts \cite{kumarage2023adv_prompt,lu2023adv_prompt}, repetition penalties \cite{fishchuk2023adv_repetition}, and human edits \cite{zhang2024mixset}. Additionally, detectors are vulnerable to adversarial attacks, including homoglyph attacks \cite{gagiano2021adv_homoglyph,macko2024ao_multilingual}, whitespace insertion \cite{cai2023adv_space}, syntactic perturbations \cite{bhat2020adv_syntactic}, synonym replacement \cite{kulkarni2023adv_semantic}, and paraphrasing \cite{krishna2024adv_paraphrasing_retrieval,shi2024adv,becker2023paraphrase}.

\section{LLM-generated Text Attribution} \label{p3}
Identifying whether a piece of text is generated by a specific LLM or a human is crucial. This distinction aids in tracing the origin of the text to ensure accountability, enhance transparency, and uphold ethical standards in information dissemination. If the content is harmful, misleading, or illegal, pinpointing the exact LLM responsible is essential for addressing ethical concerns and fulfilling legal obligations. LLM-generated text attribution builds upon techniques for LLM-generated text detection. Variations in model architecture (such as the number of layers and parameters), training methods (including pre-training and fine-tuning), and generation techniques (like sampling parameters) all subtly influence the characteristics of the generated texts \cite{munir2021aa_machine}.

\subsection{Problem Definition}
This attribution task extends beyond binary classification to handle multiple classes, increasing the complexity of LLM-generated text detection. The primary goal is to determine whether a given piece of text was created by a specific human or by one of several LLMs \cite{beigi2024model,uchendu2020aa_neural_liwc,venkatraman2023gptwho,chen2023aa_OpenLLMText,he2023mgtbench,soto2024aa_style}. A sub-problem is to attribute the text solely to LLMs, also known as model sourcing \cite{yang2023dna_gpt} or origin tracing \cite{li2023sniffer_origin_tracing}.

\subsection{Methodologies}
Attributing texts to LLMs versus human writers involves recognizing inherent differences in their text generation capabilities. LLMs typically exhibit less diversity in word usage compared to humans \cite{ippolito2019detect_bert,dugan2023eval_human}. LLMs can mimic a range of styles and tones, often masking their underlying characteristics. This ability to adapt makes attribution challenging, especially as LLMs rapidly evolve and their outputs change significantly over time \cite{guo2023data_hc3}. Additionally, LLMs may inadvertently reproduce snippets of their training data.

To simplify the classification process, it is common practice to group different human writers into a single category because humans exhibit a broader spectrum of writing styles and proficiency levels compared to machines \cite{uchendu2021bench_turing}. For example, classifications might include comparisons such as Human vs. ChatGPT, or Human vs. LLama \cite{uchendu2021bench_turing,he2023mgtbench}. Some studies have formulated a 7-class classification that includes one human class and six LLM classes \cite{he2023mgtbench,wang2024m4gt}. Other approaches consider multiple human classes, albeit with a limited number. For instance, a 10-class classification might include seven human classes and three LLM classes \cite{tripto2023hansen}. This multiclass classification is often converted into a one-vs-rest classification for each label. Transformer-based models, such as BERT and RoBERTa, are fine-tuned on datasets containing both human-written and LLM-generated texts to conduct the attribution.

\subsection{Open Challenges}
Attributing texts generated by LLMs to specific humans or models presents a multi-class classification challenge. Variations in training data, model architecture, and fine-tuning processes contribute to the distinctive outputs of different LLMs, though these differences are often subtle and difficult to detect \cite{uchendu2021bench_turing}. Effective identification requires sophisticated methods to discern the unique signatures embedded in syntactic structures and lexical choices, which are influenced by specific training datasets. However, the proprietary nature of many LLMs restricts access to comparative data, posing significant hurdles. Additionally, the high degree of stylistic overlap among LLMs, especially those with similar architectures or trained on overlapping datasets, further complicates accurate classification. Continuous updates and fine-tuning of LLMs necessitate ongoing adjustments to attribution methodologies to account for evolving model characteristics \cite{wu2023survey_llm_text}.

\section{Human-LLM Co-authored Text Attribution} \label{p4}
Besides creating text from scratch, LLMs are often used for extending sequences from human prompts. These perturbations have diminished the effectiveness of existing text detection methods \cite{bhat2020adv_syntactic}. Identifying text that combines input from both human authors and LLMs presents unique challenges. Hybrid texts may originate as human-written content, with LLMs employed for conditional generation, making it difficult to clearly distinguish between the stylistic features of human and machine contributions. There are fewer studies on this task due to its difficulty, and existing research often makes simplifications.

\subsection{Problem Definition}
A human-LLM co-authored text, also known as mixed text \cite{zhang2024mixset} or human-AI collaborative writing \cite{richburg2024coauthor3class}, is a piece of writing that is partially created by a human and then revised or extended by LLMs, and vice versa. This task involves recognizing the nuances of multi-source authorship with fine-grained precision. Some studies categorize any text that is generated, modified, or extended by a machine as human-LLM co-authored text, treating it as LLM-generated. This simplifies the task to either LLM-generated Text Detection or LLM-generated Text Attribution \cite{yang2023survey_llm_text,crothers2023survey_machine_text}. Other researchers handle LLM-revised human texts and human-revised LLM texts as a single category, alongside purely human-written and purely LLM-generated texts, approaching human-LLM co-authored text authorship attribution as a three-class classification problem \cite{zhang2024mixset,richburg2024coauthor3class}. One variation of this task is to detect the boundary between human-written and LLM-generated text \cite{cutler2021human_machine_boundary,wang2024m4gt}.

\subsection{Methodologies}

Texts generated by LLMs are generally less emotional and more objective, often repeating terms within a paragraph. In contrast, human-authored texts tend to be more coherent and exhibit greater lexical diversity \cite{guo2023data_hc3,dugan2023eval_human,zhang2024mixset}. Models like DNA-GPT \cite{yang2023dna_gpt} and DetectGPT \cite{mitchell2023detect_gpt} utilize the T5 model \cite{raffel2020t5} to simulate scenarios where humans modify LLM-generated texts. MIXSET \cite{zhang2024mixset} offers a more comprehensive dataset that includes text refined by LLMs through polishing, completion, and rewriting operations. 

To effectively analyze and classify human-LLM co-authored texts versus those solely authored by humans or LLMs, feature-based methods from LLM-generated text detection, such as Log-likelihood \cite{solaiman2019tf_idf_logistic}, GLTR \cite{gehrmann2019machine_text_statisc}, and log-rank \cite{mitchell2023detect_gpt}, are adapted to this task. Additionally, neural network-based models like BERT \cite{ippolito2019detect_bert}, Radar \cite{hu2023radar_vicuna}, and GPT-sentinel \cite{chen2023gpt_sentinel} can also be applied. The complexity of this task increases as users may employ multiple LLMs to compose different sections of an article, further blurring the lines between human and machine-generated content. Consequently, techniques used in earlier LLM detection need to evolve continuously. This ongoing evolution in detection strategies mirrors the increasing sophistication of LLM outputs and the collaborative nature of modern text creation.

\subsection{Open Challenges}
Authorship attribution involving human-written, LLM-generated, and human-LLM co-authored pieces presents varying degrees of complexity, requiring distinct analytical approaches to accurately identify and differentiate the contributions of each author. For texts authored entirely by humans or LLMs, stylometric techniques can be effectively utilized. Human-authored texts often feature unique stylistic nuances, such as variable sentence structures and emotive language \cite{zhang2024mixset}. In contrast, LLM-generated texts typically exhibit consistent syntax and a broader vocabulary \cite{guo2023data_hc3}. Feature-based methods used in LLM-generated text detection can be adapted to classify texts by identifying these distinct patterns, thereby attributing texts to their correct source.

Analyzing and classifying texts co-authored by humans and LLMs presents a significant challenge due to the blending of human and machine stylistic features. These texts often start as human drafts and are later extended or revised by LLMs, or the process might occur in reverse. This integration of styles creates a hybrid form that makes it difficult to distinguish distinct authorial markers, thereby complicating the attribution process.

Human-LLM co-authored texts pose a more intricate challenge due to the blending of stylistic and linguistic elements from both human authors and LLMs. These texts may begin as human drafts later extended or revised by LLMs, or vice versa, resulting in an integration of styles that obscures distinct authorial markers \cite{liu2024chatgpt_academic_gpabenchmark}. Advanced techniques are required to dissect these integrations, identifying where and how LLM contributions intersect with human input \cite{wang2024m4gt}. This involves detecting subtle shifts in style and contextual cues that indicate the extent of LLM involvement, allowing for accurate segmentation and attribution of authorship within hybrid documents.

\section{Resources and Evaluation Metrics} \label{resources}
This section provides an in-depth examination of widely used benchmarks and datasets in authorship attribution research, along with guidelines for selecting appropriate ones. These resources range from purely human-written texts to those generated by LLMs and human-LLM co-authored texts. This diversity is crucial for training and evaluating models across various tasks.

Traditional datasets focus exclusively on texts written by humans, while modern datasets include LLM-generated text, addressing the need to detect and attribute text produced by LLMs. Additionally, this section covers commercial and open-source detectors commonly used to identify machine-generated text. Lastly, we summarize the common evaluation metrics employed in this field.

\newcolumntype{L}[1]{>{\raggedright\arraybackslash}m{#1}}
\newcolumntype{C}[1]{>{\centering\arraybackslash}m{#1}}
\newcolumntype{R}[1]{>{\raggedleft\arraybackslash}p{#1}}

\vspace{-4mm}
\begin{table*}
\small 
\centering
\begin{tabularx}{\textwidth}{L{2.4cm}L{2.9cm}lL{1.8cm}L{1.5cm}L{3.4cm}C{0.2cm}C{0.2cm}C{0.2cm}}
    \toprule
    Name            & Domain                                                                                                            & Size               & Length                                & Language                                   & Model  & P2    & P3    & P4   \\
    \midrule
    \rowcolor[rgb]{0.753,0.902,0.961} TuringBench \newline\cite{uchendu2021bench_turing}     & News                                             & 168,612 (5.2\%)    & 100 to 400 words                      & en                                         & GPT-1,2,3, GROVER, CTRL, XLM, XLNET, FAIR, TRANSFORMER-XL, PPLM                      & \checkmark    & \checkmark    &             \\
    TweepFake \newline\cite{fagni2021tweepfake}       & Social media                                                                            & 25,572 (50.0\%)    & less than 280 characters              & en                                         & GPT-2, RNN, Markov, LSTM, CharRNN                                                    & \checkmark    &               &             \\
    \rowcolor[rgb]{0.753,0.902,0.961} ArguGPT \newline\cite{liu2023argugpt}        & Academic essays                                                                              & 8,153 (49.5\%)     & 300 words on average                  & en                                         & GPT2-Xl, text-babbage-001, text-curie-001, davinci-001,002,003, GPT-3.5-Turbo        & \checkmark    &               &             \\
    AuTexTification \newline\cite{sarvazyan2023AuText2023} & Tweets, reviews, news, legal, and how-to articles & 163,306 (42.5\%)   & 20 to 100 tokens                     & en, es                                     & BLOOM, GPT-3                                                                         & \checkmark    & \checkmark    &             \\
    \rowcolor[rgb]{0.753,0.902,0.961} CHEAT \newline\cite{yu2023data_cheat}           & Academic paper abstracts                                                                  & 50,699 (30.4\%)    & 163.9 words on average                & en                                         & ChatGPT                                                                              & \checkmark    &               &             \\
    GPABench2 \newline\cite{liu2024chatgpt_academic_gpabenchmark}       & Academic paper abstracts                                & 2.385M (6.3\%)     & 70 to 350 words                       & en                                         & ChatGPT                                                                              & \checkmark    &               & \checkmark  \\
    \rowcolor[rgb]{0.753,0.902,0.961} Ghostbuster \newline\cite{verma2023ghostbuster}     & News, student essays, creative writing                                                & 23,091 (87.0\%)    & 77 to 559 (median words per document) & en                                         & ChatGPT, Claude                                                                      & \checkmark    &               &             \\
    HC3 \newline\cite{guo2023data_hc3}             & Reddit, Wikipedia, medicine, finance                     & 125,230 (64.5\%)   & 25 to 254 words                       & en, zh                                     & ChatGPT                                                                              & \checkmark    &               &             \\
    \rowcolor[rgb]{0.753,0.902,0.961} HC3 Plus \newline\cite{su2023hc3_plus}        & News, social media                                                                          & 214,498            & N/A                                   & en, zh                                     & ChatGPT                                                                              & \checkmark    &               &             \\
    HC-Var \newline\cite{xu2023HC-Var}          & News, reviews, essays, QA                                   & 144k (68.8\%)      & 50 to 200 words                       & en                                         & ChatGPT                                                                              & \checkmark    &               &             \\
    \rowcolor[rgb]{0.753,0.902,0.961} HANSEN \newline\cite{tripto2023hansen}          & Transcripts of speech (spoken text), statements (written text)                            & 535k (96.1\%)      & less than 1k tokens                   & en                                         & ChatGPT, PaLM2, Vicuna-13B                                                           & \checkmark    & \checkmark    &             \\
    M4 \newline\cite{wang2023bench_m4}  & Wikipedia, WikiHow, Reddit, QA, news, paper abstracts, peer reviews & 147,895 (24.2\%)   & more than 1k characters               & ar, bg, en, id, ru, ur, zh                 & davinci-003, ChatGPT, GPT-4, Cohere, Dolly2, BLOOMz                                  & \checkmark    &               &             \\
    \rowcolor[rgb]{0.753,0.902,0.961} MGTBench \newline\cite{he2023mgtbench}        & News, student essays, creative writing                                                      & 21k (14.3\%)       & 1 to 500 words                        & en                                         & ChatGPT, ChatGLM, Dolly, GPT4All, StableLM, Claude                                   & \checkmark    & \checkmark    &             \\
    MULTITuDE \newline\cite{macko2023multitude_multilingual}       & News                                     & 74,081 (10.8\%)    & 200 to 512 tokens                     & ar, ca, cs, de, en, es, nl, pt, ru, uk, zh & GPT-3,4, ChatGPT, Llama-65B, Alpaca-LoRa-30B, Vicuna-13B, OPT-66B, OPT-IML-Max-1.3B  & \checkmark    &               &             \\
    \rowcolor[rgb]{0.753,0.902,0.961} OpenGPTText \newline\cite{chen2023gpt_sentinel}     & OpenWebText                                                                           & 58,790 (50.0\%)    & less than 2k words                    & en                                         & ChatGPT                                                                              & \checkmark    &               &             \\
    OpenLLMText \newline\cite{chen2023aa_OpenLLMText}     & OpenWebText                                       & 344,530 (20\%)     & 512 tokens                            & en                                         & ChatGPT, PaLM, Llama, GPT2-XL                                                        & \checkmark    & \checkmark    &             \\
    \rowcolor[rgb]{0.753,0.902,0.961} Scientic Paper \newline\cite{mosca2023bench_academia}  & Scientific papers                                & 29k (55.2\%)       & 900 tokens on average                 & en                                         & SCIgen, GPT-2,3, ChatGPT, Galactica                                                  & \checkmark    &               &             \\
    RAID \newline\cite{dugan2024raid}            & News, Wikipedia, paper abstracts, recipes, Reddit, poems, book summaries, movie reviews      & 523,985 (2.9\%)    & 323 tokens on average                 & cs, de, en                                 & GPT-2,3,4, ChatGPT, Mistral-7B, MPT-30B, Llama2-70B, Cohere command and chat         & \checkmark    &               &             \\
    \rowcolor[rgb]{0.753,0.902,0.961} M4GT-Bench \newline\cite{wang2024m4gt}     & Wikipedia, Wikihow, Reddit, arXiv abstracts, academic paper reviews, student essays & 5,368,998 (96.6\%) & more than 50 characters & ar, bg, de, en, id, it, ru, ur, zh & ChatGPT, davinci-003, GPT-4, Cohere, Dolly-v2, BLOOMz                               & \checkmark    & \checkmark    & \checkmark  \\
    MAGE \newline\cite{li2024detect_mage}           & Reddit, reviews, news, QA, story writing, Wikipedia, academic paper abstracts            & 448,459 (34.4\%)    & 263 words on average                  & en                                         & GPT, Llama, GLM-130B, FLAN-T5 OPT, T0, BLOOM-7B1, GPT-J-6B, GPT-NeoX-2               & \checkmark    &               &             \\
    \rowcolor[rgb]{0.753,0.902,0.961} MIXSET \newline\cite{zhang2024mixset}         & Email, news, game reviews, academic paper abstracts, speeches, blogs & 3.6k (16.7\%)      & 50 to 250 words            & en                                         & GPT-4, Llama2                                                                        & \checkmark    &               & \checkmark  \\
    \bottomrule
\end{tabularx}

\caption{Summary of Authorship Attribution Datasets and Benchmarks with LLM-Generated Text. Size is shown as the sum of LLM-generated and human-written texts (with the percentage of human-written texts in parentheses). Language is displayed using the two-letter ISO 639 abbreviation. Columns P2, P3, and P4 indicate whether the dataset supports problems described in Problem 2, 3, and 4, respectively.}
\label{table:benchmarks}
\end{table*}

\begin{table*}[t]
\small
\centering
\begin{tabularx}{\textwidth}{lL{2cm}L{3cm}c c X}
\toprule
Detector & Free Tier & Paid Plan & API & Humanizer & Website \\
\midrule

\rowcolor[rgb]{0.753,0.902,0.961}
GPTZero & 10k words/mo & \$12.99/mo $\cdot$ 300k words & \checkmark &  & \url{https://gptzero.me/} \\

Winston & 2k words trial & \$10/mo $\cdot$ 80k words & \checkmark &  & \url{https://gowinston.ai/} \\

\rowcolor[rgb]{0.753,0.902,0.961}
Sapling & 2k chars & \$12 $\cdot$ 100k chars & \checkmark &  & \url{https://sapling.ai/ai-content-detector} \\

Pangram & 4 checks/day & \$20/mo $\cdot$ 600 checks & \checkmark &  & \url{https://www.pangram.com/} \\

\rowcolor[rgb]{0.753,0.902,0.961}
ZeroGPT & 15k chars & \$7.99 $\cdot$ 100k chars & \checkmark & \checkmark & \url{https://www.zerogpt.com/} \\

Phrasly & 6k words & \$10.99/mo $\cdot$ unlimited & \checkmark & \checkmark & \url{https://phrasly.ai/ai-detector} \\

\rowcolor[rgb]{0.753,0.902,0.961}
Smodin AI Detector & 50k chars & \$12/mo & \checkmark & \checkmark & \url{https://smodin.io/ai-content-detector} \\

Scribbr & 500 words/check & \$19.95 $\cdot$ unlimited &  & \checkmark & \url{https://www.scribbr.com/ai-detector/} \\

\rowcolor[rgb]{0.753,0.902,0.961}
QuillBot & 1,200 words/scan & \$8.33 $\cdot$ unlimited &  & \checkmark & \url{https://quillbot.com/ai-content-detector} \\

Draft \& Goal & 2k words & \$9.99/mo $\cdot$ 200k words &  & \checkmark & \url{https://detector.dng.ai/} \\

\rowcolor[rgb]{0.753,0.902,0.961}
BrandWell & 2,500 chars & \$199/yr $\cdot$ WriteWell plan &  & \checkmark & \url{https://brandwell.ai/ai-content-detector/} \\

Undetectable AI & --- & \$5/mo $\cdot$ 10k words & \checkmark & \checkmark & \url{https://undetectable.ai/} \\

\rowcolor[rgb]{0.753,0.902,0.961}
Isgen & --- & \$8/mo $\cdot$ 350k words & \checkmark & \checkmark & \url{https://isgen.ai/} \\

Grammarly & --- & \$12/mo $\cdot$ Pro plan &  & \checkmark & \url{https://www.grammarly.com/ai-detector} \\

\rowcolor[rgb]{0.753,0.902,0.961}
Plag.AI & --- & \$14.95/mo $\cdot$ 10k words &  & \checkmark & \url{https://plag.ai/} \\

Plagiatkontroll & --- & \$15.33/mo $\cdot$ 25k words & \checkmark &  & \url{https://plagiatkontroll.no/} \\

\rowcolor[rgb]{0.753,0.902,0.961}
Originality.AI & --- & \$12.95/mo $\cdot$ 200k words & \checkmark &  & \url{https://originality.ai/} \\

CopyLeaks & --- & \$13.99/mo $\cdot$ 300k words & \checkmark &  & \url{https://copyleaks.com/ai-content-detector} \\

\rowcolor[rgb]{0.753,0.902,0.961}
GPT Radar & --- & \$0.02 $\cdot$ 100 tokens &  &  & \url{https://gptradar.com/} \\

Turnitin's AI detector & --- & License required &  &  & \url{https://www.turnitin.com/} \\

\bottomrule
\end{tabularx}
\caption{Overview of commercial LLM-generated text detectors.}
\label{table:commercial_detectors}
\end{table*}

\begin{table*}[h!]
\small
\centering
\begin{tabularx}{\textwidth}{lL{5.5cm}X}
\toprule
Detector & Method & Repository \\
\midrule

\rowcolor[rgb]{0.753,0.902,0.961}
Binoculars & Zero-shot & \url{https://github.com/ahans30/Binoculars} \\

DetectGPT & Zero-shot $\cdot$ probability curvature & \url{https://github.com/eric-mitchell/detect-gpt} \\

\rowcolor[rgb]{0.753,0.902,0.961}
Fast-DetectGPT & Zero-shot $\cdot$ conditional probability curvature & \url{https://github.com/baoguangsheng/fast-detect-gpt} \\

GPT-2 Output Detector & RoBERTa fine-tune & \url{https://github.com/openai/gpt-2-output-dataset/tree/master/detector} \\

\rowcolor[rgb]{0.753,0.902,0.961}
Hello-SimpleAI Detector & RoBERTa trained on HC3 & \url{https://huggingface.co/Hello-SimpleAI/chatgpt-detector-roberta} \\

Desklib AI Text Detector & DeBERTa-v3 trained on RAID & \url{https://huggingface.co/desklib/ai-text-detector-v1.01} \\

\bottomrule
\end{tabularx}

\caption{Overview of open-source LLM-generated text detectors.}
\label{table:opensource_detectors}
\end{table*}

\subsection{Benchmarks and Datasets}
Authorship datasets encompass a wide range of sources, from formal literature to informal online communications, highlighting the increasing significance of user-generated content on social media. Human authorship datasets typically include author identifiers and ideally contain multiple texts for each author. Manually collecting data for large datasets is time-consuming and costly, motivating researchers to utilize web data sources such as Wikipedia and Reddit. In contrast, custom datasets for LLM-generated text are easier and more affordable to create, and they are often built alongside human-written text to maintain a similar domain and format. 

A general guideline for selecting and constructing datasets involves incorporating variations in domain, model architecture, and decoding strategies. Addressing class imbalance is crucial, as LLM-generated and human-written texts are often disproportionate. For human authorship data, it is recommended to choose content created before the widespread use of LLMs (GPT-3 \cite{brown2020gpt3} was released in June 2020, and ChatGPT followed in November 2022) to ensure that the texts were predominantly human-written.

Factors influencing the performance of existing authorship attribution models include the size of the training text \cite{hirst2007bigram_syntax,marton2005compression_text_classify}, the number of candidate authors \cite{koppel2006aa_meta_learning}, and the imbalanced distribution of training texts among the candidate authors \cite{stamatatos2008aa_sampling_imbalance}. The availability of digital text in formats such as tweets, blogs, and articles has exponentially increased, providing more training data to accelerate the development of authorship attribution. However, the rapid growth of online communication has also led to shifts in writing behavior, resulting in shorter, fragmented, and less coherent social media posts and text messages. For example, tweets are limited to 280 characters, whereas legal judgment documents contain thousands of words \cite{seroussi2011legal_judgment}. The challenge in social media stems from the brief nature of posts and a large pool of potential authors, making the attribution of short documents particularly difficult \cite{aborisade2018tweet_logistic_regression,seroussi2014data_imdb_aa_topic_model,theophilo2021social_msg_short}.

Table \ref{table:benchmarks} provides a comprehensive overview of widely used benchmarks and datasets. These datasets are characterized by statistics including domain, size, word length, language, and the LLMs used to generate texts. All listed datasets support LLM-generated text detection (Problem 2). However, fewer support LLM-generated text attribution (Problem 3) and Human-LLM Co-authored Text Attribution (Problem 4). These benchmarks often originate from human-written datasets like XSum \cite{narayan2018xsum}, OpenWebText \cite{Gokaslan2019OpenWebText}, and Wikipedia. Since Problems 2, 3, and 4 frequently treat human-written text as a single category for simplicity, rather than identifying individual authors as in Problem 1, many of these datasets are unsuitable for Problem 1, which requires unique author identification. Therefore, the representative datasets for human authorship are summarized as follows:

\begin{itemize}
    \item {Amazon Review} \cite{ni2019data_amazon}: Featuring reviews with ratings, text, votes, product metadata, and links, this dataset provides a comprehensive view of consumer opinions, ideal for commercial authorship attribution studies.
    
    \item {Aston 100 Idiolects Corpus} \cite{tian2023aston100}: Comprising emails, essays, text messages, and business memos from 100 individuals (ages 18–22, native English speakers), this corpus provides a broad spectrum of text types for analyzing both content and stylistic features.
    
    \item {Blog Authorship Corpus} \cite{schler2006data_blog}: Contains over 680,000 posts from more than 19,000 authors, with an average of 35 posts per author. The texts from the dataset are informal and conversational, typical of blog posts.

    \item {Deceptive Opinion Spam} \cite{ott2011data_opinion_spam}: Includes 400 genuine and 400 deceptive hotel reviews, with deceptive reviews generated using Amazon Mechanical Turk, useful for studying the nuances of fake versus real reviews.  %

    \item {Enron Email} \cite{klimt2004data_enron}: Includes around 500,000 messages from 160 employees, offering long texts and high text-per-author variance, making it ideal for studying corporate communication styles.
    
    \item {Fanfiction}: Collected from fanfiction.net, this dataset includes fan-written fiction \cite{bischoff2020data_fanfiction,kestemont2021survey_pan21}, providing insights into creative writing and authorship attribution in fictional narratives.
        
    \item {IMDb1M} \cite{seroussi2014data_imdb_aa_topic_model}: Features over 270,000 movie reviews by 22,000 authors, with an average of 12.3 texts per author and an average text length of 121 tokens, suitable for analyzing shorter, user-generated content.

    \item {Pushshift Reddit} \cite{baumgartner2020data_reddit}: This dataset comprises posts and comments from various subreddits, covering diverse topics and writing styles, making it suitable for analyzing informal online discourse.
    
    \item {PAN} \cite{kestemont2021survey_pan21,bevendorff2022pan}: Offered by PAN workshops for benchmarking authorship attribution and verification models and are used in various authorship attribution competitions.

    \item {VALLA} \cite{tyo2022bench_valla}: Designed for benchmarking authorship attribution models, VALLA standardizes a range of texts across various genres and writing styles.
\end{itemize}

Tables \ref{table:commercial_detectors} and \ref{table:opensource_detectors} summarize representative commercial and open-source LLM-generated text detectors. Commercial systems are mainly designed for LLM-generated text detection (Problem 2), while some, such as GPTZero \cite{tian2023gptzero}, also support human-LLM co-authorship analysis and boundary detection, partially addressing Problem 4 \cite{cutler2021human_machine_boundary}. Recently, many commercial platforms have also integrated ``Humanizer'' functions, reflecting an emerging detector-evasion ecosystem in which detection and text rewriting tools co-evolve. Although these systems often advertise very high accuracy, few are evaluated on shared benchmarks, and existing studies show persistent issues with false positives, domain shift, sampling variation, adversarial rewriting, and unseen language models \cite{dugan2024raid}. Open-source detectors provide more transparent baselines, but they similarly face robustness and generalization challenges.

\subsection{Evaluation Metrics}

Evaluation metrics such as F1 score and AUCROC are essential for quantifying the performance of authorship models, providing a standardized means to assess and compare the effectiveness of different authorship attribution approaches. As in other classification tasks, existing studies predominantly use the Area Under the Receiver Operating Characteristic (AUCROC) and F1 score to evaluate attribution algorithms. In human authorship attribution, where there are a large number of candidate authors, retrieval metrics like Mean Reciprocal Rank (MRR) and recall-at-k are used \cite{rivera2021luar}. Additionally, Self-BLEU and perplexity are useful metrics, with a lower Self-BLEU score indicating higher textual diversity \cite{zhang2024mixset}. Common evaluation metrics include:

\begin{itemize}
    \item Accuracy: Measures the proportion of correctly identified authors but can be misleading in imbalanced datasets.

    \item Precision, Recall, and F1-Score: Crucial in imbalanced datasets. Precision indicates the relevance of identified instances, recall measures the ability to identify all relevant instances, and the F1-Score balances both.

    \item Area Under the Receiver Operating Characteristic Curve (AUCROC): Represents the trade-off between true positive rates and false positive rates, where higher values indicate better performance.

    \item False Positive Rate (FPR) and False Negative Rate (FNR): Critical for minimizing misclassification, with FPR measuring incorrect classification of human texts as LLM-generated and FNR the reverse.

    \item Recall-at-k: Measures the probability that the correct author appears among the top k results when ranking targets by cosine similarity to a query text.

    \item Mean Absolute Error (MAE): Used to evaluate the performance of human-machine text boundary detection. It measures the average absolute difference between the predicted position index and the actual change point.

\end{itemize}

\section{Opportunities and Future Directions} \label{opportunities}
This section explores future directions in the field of authorship attribution, focusing on leveraging the potential of LLMs while addressing associated challenges. Future efforts should aim for finer granularity in authorship attribution, leveraging LLM capabilities, improving generalization, enhancing explainability, preventing misuse, developing standardized benchmarks, and integrating interdisciplinary perspectives to enrich the field.

\subsection{Finer Granularity}
Current authorship attribution methods often face limitations when handling a more extensive range of candidate human authors or LLMs, presenting opportunities for future research. For instance, existing approaches for LLM-generated Text Attribution typically manage only a limited number of authors or models, which restricts their applicability in real-world scenarios where the pool of potential authors or models can be vast. Previous studies have often oversimplified the problem by categorizing all human-written text as a single category \cite{uchendu2021bench_turing,he2023mgtbench}. This approach ignores the diversity among human authors and fails to leverage the rich set of characteristics that distinguish individual writing styles. Future work can build upon traditional research on human authorship to develop methods capable of attributing human-written text to individual authors even within the context of LLM-generated content. This refinement will improve the accuracy and utility of authorship attribution models, especially in mixed datasets containing both human-written and LLM-generated texts.

Similarly, for Human-LLM Co-authored Text Attribution, there is a need to attribute text more precisely to individual human authors or specific LLMs. Current work simplifies human-written text, LLM-generated content, and text co-authored by humans and LLMs into three broad categories, without differentiating between individual human authors and specific LLMs \cite{zhang2024mixset,richburg2024coauthor3class}. This approach overlooks the nuanced contributions of each author or model. By improving the granularity of attribution, future models can better distinguish between various human authors and LLMs, thus increasing the practicality and reliability of authorship attribution tools. Such advancements would be particularly valuable in collaborative environments where multiple human authors and LLMs contribute to a single body of work, enabling clearer recognition of each contributor's role.

\subsection{Generalization}
This subsection examines the applicability of current methodologies across varying LLMs, domains, genres, and languages. Domains refer to broad areas of knowledge or topics, while genres refer to specific styles or forms of writing within any domain. Domain generalization poses significant challenges due to variability in vocabularies, syntax, and styles across different subjects, complicating accurate authorship attribution. Attribution performance tends to drop when known and query texts differ in topic or genre \cite{altakrori2021cross_topic}. Models like BERT and RoBERTa have shown limitations in cross-domain tasks \cite{barlas2020aa_cross_domain_bert,huertas2022part}, and adapting models to new domains remains difficult due to factors like dataset size variability and writing styles. Traditional methods focused on identifying less topic-dependent features such as function words and part-of-speech n-grams \cite{madigan2005cross_ngrams,menon2011cross_ngrams}, while recent approaches highlight the importance of training more powerful transformer-based models \cite{rivera2021luar} and techniques such as adversarial training \cite{li2017cross_adv_senti_class,ben2010adv_theory,ganin2016domain_adv}.

Genre generalization involves adapting to different writing styles, such as fiction, nonfiction, and poetry, each with unique features. Authors' adaptation to various genres dilutes their identifiable stylistic traits, complicating attribution. Similarly, distinguishing between human-written and LLM-generated text in different genres requires identifying genre-specific inconsistencies. The diversity of genres demands flexible models capable of understanding various narrative structures, tones, and stylistic elements. Adapting models to handle genre variations necessitates more advanced and flexible approaches for effective generalization. Current authorship attribution models also struggle with out-of-distribution issues when faced with languages and LLMs not encountered during training, leading to decreased accuracy and reliability \cite{koppel2005style_native_language,wu2023survey_llm_text}. Addressing this generalization problem is crucial for developing robust models that can handle diverse and evolving linguistic and model landscapes.

Improving generalization can potentially be achieved through several strategies. First, leveraging transfer learning by pre-training on large, diverse datasets and fine-tuning on specific domains enhances adaptability and performance \cite{barlas2020aa_cross_domain_bert,rodriguez2022cross_academia}. Second, developing domain and genre-invariant features would improve robustness by focusing on core stylistic elements \cite{argamon2003writing_style_gender}. Third, employing hybrid models and ensemble methods that integrate domain-specific knowledge can optimize prediction accuracy by drawing on the strengths of individual models \cite{bacciu2019aa_cross_domain_multi_lang_ensemble_svm}. Additionally, incorporating contextual factors like the writing environment or intended audience, alongside data augmentation techniques, can bolster generalization. Finally, improving attribution in multilingual contexts enables models to operate effectively across various languages \cite{chen2022chinese,shamardina2022russian}. As detectors could be biased against non-native English writers \cite{liang2023bias_non_en}, enhancing multilingual generalization is crucial for fairness. Collectively, these approaches foster robust and adaptable models equipped to handle diverse styles and contexts.

\subsection{Explainability}
Improving explainability is crucial for ensuring transparency and trust in authorship attribution models, especially as they become increasingly integrated into fields such as forensic investigation, academia, and journalism. Developing explainable AI techniques specifically for authorship attribution can lead to methodologies where the reasoning behind attributions is clear and understandable. This transparency is particularly important when authorship attribution is used as evidence in legal proceedings \cite{chaski2005forensic_court,rocha2016aa_law}. Traditional attribution methods, which rely on stylistic and linguistic features to identify an author, often struggle to distinguish between human-authored texts and those generated by LLMs, which can adeptly replicate these features. This challenge underscores the need for enhanced methodologies that not only differentiate between human and AI-generated content but also explore how LLMs emulate specific authorial styles \cite{boenninghoff2019explain_siamese_lstm,danilevsky2020explain_nlp}.

Despite attempts such as analyzing internal attention weights or employing interpretation tools and visualization techniques \cite{wallace2019explain_allennlp}, word-level explanations are insufficient. The challenge remains to provide higher-level explainability that aligns with human cognitive processes \cite{rudin2019explain_black_box}. Advances in this area may include leveraging discourse-level relations and training models with human explanations for common sense reasoning to improve the explanatory depth of model-generated attributions \cite{rajani2019explain_nlp_commonsense}. For instance, \citet{kowalczyk2022review_shap} detected GPT-2-generated fake reviews using Shapley Additive Explanations (SHAP) \cite{lundberg2017shap}.

\vspace{-0.5\baselineskip}
\subsection{Misuse Prevention}
Future research should focus on refining existing authorship attribution methods to detect and prevent malicious activities such as generating misinformation, plagiarism, and propaganda \cite{goldstein2023generative_model_misuse,hazell2023phising,spitale2023disinfomation,lund2023chatgpt_academic}. These methods analyze stylistic features to detect discrepancies in claimed authorship and trace the origins of content, thereby identifying suspicious texts. For plagiarism, attribution models can compare writing styles to a database of known authors. In combating misinformation and propaganda, these models could identify and flag content patterns typical of known propagandists.

To ensure effectiveness in real-world tasks, authorship attribution models should be robust against out-of-domain data and adversarial attacks. Adversarial attacks, such as alternative spellings, article deletion, paragraph additions, case changes, zero-width spaces, whitespace manipulation, homoglyphs, number swaps, misspellings, paraphrasing, and synonym substitution, have been shown to effectively degrade detector performance \cite{dugan2024raid}. The emergence of commercial ``Humanizer'' tools as shown in Table \ref{table:commercial_detectors} further intensifies this challenge by explicitly rewriting LLM-generated text to appear more human-written, creating a detector-evasion feedback loop. Future attribution systems should therefore evaluate robustness not only against simple perturbations, but also against humanizer-assisted paraphrasing and style transfer. Diversifying training data with various writing styles and topics could improve out-of-domain robustness. Adversarial robustness could be achieved through adversarial training and employing ensemble methods to build resilience against intentional manipulations.

\subsection{Leveraging LLM Capabilities}
Leveraging LLMs can enhance both traditional feature-based stylometry methods and LLM-based approaches. By integrating LLMs with existing methods, researchers can gain deeper insights into stylistic nuances, improving the robustness of authorship detection across various textual genres and lengths. The increasingly large context length of LLMs enables in-context learning (ICL) \cite{brown2020gpt3} by incorporating more documents, enhancing the model's ability to capture intricate writing patterns.

Another promising approach is Retrieval-Augmented Generation (RAG) \cite{lewis2020rag}. RAG can enhance authorship attribution by retrieving additional documents for each author, thereby assisting in generating more contextually accurate results. Moreover, leveraging LLMs for data augmentation and synthetic data generation can create diverse training datasets, which in turn improves the generalization of attribution models \cite{albalak2024data_selection}.

Combining text detection with other modalities, such as images and videos, can potentially enhance authorship attribution. Multi-modal analysis offers a more comprehensive understanding of the content and its origins by integrating diverse data types. This holistic approach not only improves the attribution process but also fosters the development of innovative methodologies that are better adapted to the evolving nature of digital content.

\subsection{Developing Standardized Benchmarks}

The diversity in datasets and evaluation metrics currently poses a challenge to the consistent and fair evaluation of different authorship attribution methods. To advance the field, it is crucial to establish benchmarks that encompass a broad spectrum of text types and sources, including those that are human-authored, LLM-generated, and human-LLM co-authored. These benchmarks should integrate a diverse range of text corpora, representing various sources, domains, text lengths, and languages, to accurately reflect real-world applications. Moreover, unified and comprehensive evaluation metrics are essential for ensuring consistent and transparent measures of attribution accuracy, robustness, explainability, and computational efficiency, thereby enabling fair comparisons across different models. Additionally, benchmarks should include datasets that combine human-written and machine-generated content from various authors and LLMs, simulating realistic tasks and testing the robustness of attribution models.

To ensure ongoing relevance and challenge for attribution methods, benchmarks must be regularly updated to include new types of LLMs and detectors. By developing and adopting standardized benchmarks, the research community can foster more rigorous, reproducible, and comparable studies. This will ultimately drive advancements in authorship attribution methodologies and applications. These standardized benchmarks would serve as a foundation for the systematic evaluation of attribution techniques, promoting innovation and progress in addressing the complexities of authorship attribution in a rapidly evolving digital landscape.

\subsection{Integrating Interdisciplinary Perspectives}
Authorship attribution is inherently multidisciplinary, encompassing elements of linguistics, computer science, forensic science, and psychology \cite{stamatatos2009survey}. Future research should continue fostering collaboration across these fields to integrate diverse perspectives and methodologies. This integrative approach can lead to innovative solutions and a deeper understanding of the challenges and potential of authorship attribution. Combining insights from various disciplines can foster the creation of holistic attribution models that account for both the intricacies of human language and the complexities of LLM-generated texts. Such collaboration could also spearhead initiatives to standardize evaluation metrics for authorship attribution tools, ensuring their effectiveness across diverse contexts and compliance with ethical standards.

Linguistics can dissect textual structures and stylistic nuances, identifying unique linguistic fingerprints of authors. It explores novel features that enhance robustness across domains and improve explainability. Forensic science contributes through technological tools and methodologies, enabling precise examination of physical and digital texts. Psychology, particularly psycholinguistics, provides insights into how the brain processes function words and grammatical markers distinctively from lexical content words, revealing correlations with socio-cultural categories such as gender, age, and native language \cite{chambers2013sociolinguistic,nerbonne2014psycho_linguistic,seals2023psycholinguistic}, which are pivotal in understanding identity and social affiliations \cite{argamon2009profiling_bayes}. The Linguistic Inquiry and Word Count (LIWC) tool exemplifies how automated text analysis can use more than 100 psychological dimensions to analyze word use, reflecting distinct language variations among different groups in specific genres and languages \cite{goldstein2009cross_liwc,pennebaker2015psychology_liwc,duduau2021liwc_social}.

Combining these interdisciplinary perspectives enhances our ability to distinguish between human and machine-generated texts, addressing the emerging challenges posed by sophisticated language models. By integrating linguistic theory with advanced computational techniques, forensic methodologies, and psychological insights, researchers can develop more comprehensive and nuanced authorship attribution frameworks. These frameworks will be better equipped to handle the diverse range of writing styles and contexts, ultimately leading to more accurate and reliable attribution outcomes. Furthermore, interdisciplinary collaboration can drive the development of ethical guidelines and best practices, ensuring that authorship attribution is conducted responsibly and with respect for individuals' privacy and rights.

\section{Ethical and Privacy Concerns} \label{ethic}
In the evolving landscape of authorship attribution, it is crucial to prioritize ethical considerations to safeguard privacy, integrity, and the rightful ownership of content. The attribution of text to specific authors or models raises significant ethical and privacy issues. Misattribution can lead to wrongful accusations or misinterpretation of an author's intent \cite{lund2023chatgpt_academic}. Additionally, the use of attribution technologies must balance the need for accountability with respect for individuals' privacy and the potential for misuse in surveilling or censoring content.

Authorship attribution techniques are essential in digital forensics, cybersecurity, and plagiarism detection. However, the potential to reveal the identities of anonymous authors presents significant ethical challenges. Applications such as linking user accounts across platforms and identifying compromised accounts raise privacy concerns and ethical questions about surveillance and profiling individuals based on their writing style.

The use of authorship attribution methods must be carefully managed to protect individual privacy and adhere to ethical standards, particularly in sensitive areas like journalism, political dissent, and corporate whistle-blowing \cite{sison2023ethic}. Ensuring that these methods are not used to undermine privacy rights or expose individuals to risks without their consent is essential. Despite existing measures to prohibit the unethical use of LLMs, these restrictions could be evaded through prompt engineering and jail-breaking, posing risks of phishing and fraud scams.

Furthermore, the increasing difficulty in distinguishing between human and LLM-generated content raises concerns about intellectual property, plagiarism, and accountability. Accurate attribution is crucial for maintaining academic and creative integrity, yet tools and methods for achieving this must evolve rapidly to keep pace with technological advancements. The deployment of LLMs in generating content across various domains, from journalism to literature, necessitates a rethinking of authorship norms and the legal frameworks governing creative works.

\section{Conclusion} \label{conclusion}
The field of authorship attribution is experiencing both unprecedented challenges and remarkable opportunities with the advent of LLMs. Whether the objective is to identify human authors, differentiate between human and machine-generated texts, attribute texts to specific LLMs, or manage the complexities of human-LLM co-authored texts, ongoing innovation is imperative. Effectively addressing these multifaceted issues requires interdisciplinary approaches and collaborative efforts among researchers. This review explores various problems within authorship attribution, offering a comprehensive comparison of methodologies and datasets. By integrating robustness, explainability, and interdisciplinary perspectives, the insights gained are not only accurate but also socially relevant and trustworthy. The review highlights the strengths and limitations of current approaches, identifies key open problems, and outlines future research directions. This holistic analysis equips researchers and practitioners with the knowledge necessary to navigate the evolving landscape of authorship attribution, emphasizing critical areas for future research and development.

\section*{Acknowledgments}
This material is based upon work supported by the U.S. Department of Homeland Security under Grant Award Number 17STQAC00001-07-04, NSF awards (SaTC-2241068, IIS-2339198, and POSE-2346158), a Cisco Research Award, and a Microsoft Accelerate Foundation Models Research Award. The views and conclusions contained in this document are those of the authors and should not be interpreted as necessarily representing the official policies, either expressed or implied, of the U.S. Department of Homeland Security and the National Science Foundation.

\bibliographystyle{ACM-Reference-Format}
\bibliography{biblio}

\end{document}